\documentclass[aps,prb,amsmath,amssymb,twocolumn,10pt,superscriptaddress]{revtex4-2}
\usepackage{graphicx}
\usepackage{dcolumn}
\usepackage{bm}
\usepackage{times}
\usepackage[colorlinks,citecolor=blue]{hyperref}
\usepackage{xcolor} 
\usepackage{kotex}
 \usepackage{ulem}
\usepackage{rotating}

\begin{document}

\preprint{APS/123-QED}

\title{Quasidegenerate charge-density wave states in 1\textit{T}-TiSe$_2$}% Force line breaks with \\

\author{Seungrok Mun}
%\email{lucas3691@kangwon.ac.kr}
\affiliation{%
 Department of Semiconductor Physics, Kangwon National University, Chuncheon 24341, Republic of Korea 
}
\affiliation{%
Interdisciplinary Program in Earth Environmental System Science and Engineering, Kangwon National University, Chuncheon 24341, Republic of Korea
}

\author{Woojin Choi}
\affiliation{%
 Department of Semiconductor Physics, Kangwon National University, Chuncheon 24341, Republic of Korea 
}
\affiliation{%
 Institute of Quantum Convergence Technology, Kangwon National University, Chuncheon 24341, Republic of Korea
}
\author{Hayoon Im}
\affiliation{%
 Department of Physics, Pusan National University, Busan, Republic of Korea 
}

\author{Sung-Kwan Mo}
\affiliation{%
 Advanced Light Source, Lawrence Berkeley National Laboratory, Berkeley, CA 94720, USA 
}

\author{Choongyu Hwang}
\affiliation{%
 Department of Physics, Pusan National University, Busan, Republic of Korea 
}

\author{Jinwoong Hwang}
\email{jwhwang@kangwon.ac.kr}
\affiliation{%
 Department of Semiconductor Physics, Kangwon National University, Chuncheon 24341, Republic of Korea 
}
\affiliation{%
 Institute of Quantum Convergence Technology, Kangwon National University, Chuncheon 24341, Republic of Korea 
}

\author{Heung-Sik Kim}
\email{heungsikim@kentech.ac.kr}
\affiliation{%
 Department of Semiconductor Physics, Kangwon National University, Chuncheon 24341, Republic of Korea 
}
\affiliation{%
 Institute of Quantum Convergence Technology, Kangwon National University, Chuncheon 24341, Republic of Korea 
}
\affiliation{%
Interdisciplinary Program in Earth Environmental System Science and Engineering, Kangwon National University, Chuncheon 24341, Republic of Korea
}
\affiliation{%
Department of Energy Engineering, Korea Institute of Energy Technology, Naju-si 58217, Republic of Korea
}

\begin{abstract} Transition metal dichalcogenides have been actively studied for their intriguing charge density wave (CDW) formations and their impacts on material properties. Among these, 1\textit{T}-TiSe$_2$ is well-known to exhibit a 2$\times$2$\times$2 CDW state transition at 200 K, but its true ground state nature remains under debate.
In this study, we investigate possible CDW states in 1\textit{T}-TiSe$_2$ and their consequences for transport properties by employing first-principles electronic structure calculations and angle-resolved photoemission spectroscopy. We identify seven distinct types of 2$\times$2$\times$2 CDW phases, most of which have not been reported previously. All of these phases are nearly degenerate in energy with each other ($< 1.41$ meV per formula unit).
Using the band unfolding technique, we compare the electronic band structures of these CDW phases with experimental angle-resolved photoemission spectroscopy data. Our findings support the presence of a possible second phase transition at 165 K and suggest a new intermediate CDW order between 165 and 200 K that was previously unexplored. This result provides a possible resolution to the conflict between previous reports on the ground state symmetry of 1\textit{T}-TiSe$_2$, and opens a viable route to phase engineering of 1\textit{T}-TiSe$_2$ for functional applications. 
\end{abstract}

\maketitle

%\tableofcontents

\section{\label{sec:level1}introduction}

Transition metal dichalcogenides (TMDC) are layered materials with the general formula MX$_2$, where M and X denote a transition metal and a chalcogen element, respectively.
TMDC have been actively studied in recent years for their fundamental properties as realizations of two-dimensional electronic and magnetic systems \cite{burch2018magnetism, manzeli20172d}, as well as for their promising characteristics for potential electronic, optoelectronic \cite{wang2012electronics}, and spintronic device applications \cite{manzeli20172d}.
Extensive studies on TMDC have demonstrated that heterostructuring different TMDC layers and tuning the layer thickness can lead to the emergence of unique electronic and optical properties. By utilizing these distinctive characteristics, researchers are actively exploring their potential applications in semiconductor technologies\cite{geim2013van,wang2012electronics,manzeli20172d}.

%{\color{violet}\bf
%(HSK: TMDC에서 스핀 액상이나 위상 절연체 이슈는 상당히 좁은 범위의 물질에 한정되어 있고, 이보다는 반도체 응용이나 optoelectronics 분야에 보다 더 많이 이용되는 듯 합니다. TMDC에 대한 보다 일반적인 소개를 위해 리뷰 논문등을 참고해 보세요: Geim and Grigorieva, Nature 499, 419–425 (2013), Wang et al., Nature Nanotechnology 7, 699–712 (2012), Manzeli et al., Nature Reviews Materials 2, 17033 (2017))
%}
%{\color{orange} 적어주신 논문 참고하여 TMDC의 heterostructure와 layer 두께에 따른 새로운 특성이 나온것과, 독특한 물리적 광학적 특성이 나오는것을 적어보았습니다.}

One of the most notable phenomena observed in TMDC is the emergence of charge density waves (CDW), which manifest as periodic modulations in the electronic and atomic distributions. The origins of CDW in these materials are attributed to mechanisms such as Fermi surface nesting, electron–phonon coupling, and exciton condensation \cite{Gruner1994,rossnagel2001fermi,song2023signatures}. Such CDW phenomena have been intensively studied for their potential relation to strong correlation effects such as superconductivity and magnetism \cite{navarro2016enhanced,manas2021quantum,cossu2024stacking}. Furthermore, layer stacking of CDW patterns has been suggested to affect the electronic properties of transition metal dichalcogenides and other layered materials, adding another degree of freedom in tuning the material properties of these quasi-two-dimensional systems \cite{yang2024origin,li2021reorganization,ritschel2018stacking,wu2022effect,cossu2024stacking}.
Among TMDC materials, 1\textit{T}-TiSe$_2$ has shown many interesting properties in its CDW phase, but the nature of many observations has remained elusive, for example, regarding the microscopic origin of the CDW formation \cite{Kusmartseva2009,Kidd2002,Suzuki1985,Cercellier2007}. Regarding the ground state nature of 1\textit{T}-TiSe$_2$, early experimental studies reported a 2$\times$2$\times$2 CDW phase at 200 K and found only a single CDW phase\cite{DiSalvo1976,Holt2001}. 
However, there has been an ongoing debate on the nature of the ground-state CDW pattern, as different experimental studies have reported distinct structural phases with different space group symmetries \cite{kim2024origin,ueda2021correlation,wickramaratne2022photoinduced,jog2023optically,hellgren2017critical}.

%
%{\color{violet}\sout {(여기 TiSe2의 구조에 대해 상반된 결과를 알려주는 실험 논문들 몇 개 cite 부탁합니다. 전에 찾아보고 알려준 논문들이 많았지요? 그리고 TiSe2의 CDW mechanism이 electron-phonon interaction인지, 아니면 excitonic insulator 형성인지에 대한 논쟁, 그리고 TiSe2의 ground state가 semiconducting인지 metallic인지에 대한 논쟁 또한 여전히 진행중입니다. 뒤 discussion에 제가 언급한 Hellgren et al., PRL 119, 176401 (2017) 논문 같은 경우는 HSE 계산 및 실험 결과를 근거로 TiSe2가 semiconducting하다고 하는데, 황진웅 교수님께서 언급하신 transport 측정 결과 및 ARPES 결과를 보면 metallic한 것처럼 보입니다. 따라서 우리 논문은 사실상 metallic 한 시나리오, 그러므로 excitonic insulator 시나리오보다는 electron-phonon coupling에 의한 CDW 형성을 지지하게 됩니다 (excitonic insulator 시나리오에서는 밴드가 여전히 metallic한 것이 부자연스럽습니다, 왜냐면 금속 상태에서는 캐리어에 의한 screening때문에 electron-hole 사이의 Coulomb potential이 강하게 억제되기 때문!). 일단 승록씨가 인용한 Hellgren PRB 2021년 논문은 이 물질이 semimetal이라 이야기하고 있기도 하고... 반면에, 버클리 그룹에서 나온 다음 논문은 CDW에 의한 semiconducting state를 지지하고 있습니다 (이 논문은 우리와 상당히 유사한 접근을 시도하고 있으니 한 번 꼭 읽어 보세요: https://www.science.org/doi/full/10.1126/sciadv.adl4481 )
%이러한 내용들을 뒤 result 및 discussion에 언급해야 하고, 이에 대한 사전 밑밥을 introduction에 깔아야 합니다. 논문 서치 및 정리, 그리고 읽어 보는 것을 좀 더 시간을 들여 잘 해 봐야 합니다.)}}
%  
Even more interestingly, recent X-ray diffraction (XRD) data and polarized Raman spectroscopy \cite{kim2024origin,ueda2021correlation}, together with our temperature-dependent resistivity measurements (see Supplementary Figure S1), have revealed a potential second CDW phase transition around 165 K, where the resistivity shows a change from semiconducting to metallic behavior as the temperature is lowered.
This observed resistivity peak has been suggested to occur due to an electron- to hole-like transport behavior \cite{King2019}, a Fermi surface reconstruction \cite{Knowles2020}, or a coherent-to-incoherent transition in the electronic dispersion \cite{Yi2024}. However, no conclusive understanding of the nature of these behaviors has been reached yet.
In this regard, one may ask whether the resistivity crossover from semiconducting to metallic behavior around 165 K, as observed in 1\textit{T}-TiSe$_2$, can be related to a possible change in the CDW stacking pattern.

In this study, we employ first-principles electronic structure calculations to investigate possible CDW patterns in the 2$\times$2$\times$2 supercell of 1\textit{T}-TiSe$_2$.
%We systematically explore various van der Waals (vdW) corrections and spin-orbit coupling (SOC) effects to obtain a comprehensive understanding of CDW phases.
%
Our calculations reveal a total of seven distinct metastable CDW phases, which exhibit extremely small energy differences of less than 1.41 meV per formula unit. This quasidegeneracy provides a way to understand the conflicting experimental reports on the ground state symmetry of this compound \cite{ueda2021correlation,kim2024origin,hellgren2017critical,wickramaratne2022photoinduced}, but also makes it challenging to clearly identify the ground state using standard density functional theory (DFT) calculations alone. To circumvent this issue, we compare the unfolded DFT band structures with angle-resolved photoemission spectroscopy (ARPES) data. From this comparison, we deduce that the ground state CDW phase is in the $P\bar{3}c1$ space group symmetry, which transforms into an intermediate $C2/m$ phase in the range of $165 < T < 200$ K. The calculated density of states shows the opening of a pseudogap feature at the Fermi level as the transition from $C2/m$ to $P\bar{3}c1$ occurs, providing an explanation for the observed resistivity peak and supporting the CDW transition scenario at 165 K.

Our results imply that, due to the small energy differences between the metastable phases, different stacking patterns that may result in lower symmetry \cite{kim2024origin} or even spatially heterogeneous CDW order \cite{wang2025} can occur.
This finding is particularly significant as it supports recent experimental proposals suggesting the existence of chiral CDW states in 1\textit{T}-TiSe$_2$ \cite{kim2024origin}.
%The absence of mirror symmetry provides strong evidence that the mixed CDW phase may be a potential candidate for the emerging chiral CDW scenario.
%
Overall, our combined theoretical and experimental results identify a series of metastable CDW phases, reconciling various conflicting observations on the nature of 1\textit{T}-TiSe$_2$, and further provide insights into its stability and symmetry properties as well as potential application of TiSe$_2$ for phase engineering and functional applications.

\section{\label{sec:level2} Methods}

The DFT calculations were performed using {\sc vasp} (Vienna Ab initio Simulation Package) \cite{kresse1993,Kresse1994JPCondMat,kresse1996efficient,Kresse1996CMS,kresse1999}. Structural relaxations were carried out using a 9$\times$9$\times$6 Monkhorst–Pack grid, with a plane-wave energy cutoff set to 400 eV. All calculations were conducted using the Perdew–Burke–Ernzerhof (PBE) generalized gradient approximation for the exchange–correlation potential \cite{perdew2008restoring}, with an additional van der Waals (vdW) energy correction. We employed the optB86b vdW functional, and additionally used the DFT-D2 and Tkatchenko–Scheffler methods, as well as r$^2$SCAN + rVV10, for further verification of the results \cite{klimevs2011van,grimme2006semiempirical,tkatchenko2009accurate,PhysRevX.6.041005}. Total energy and force convergence criteria of 10$^{-8}$ eV and 10$^{-4}$ eV/\AA were adopted for the charge self-consistent calculations and lattice optimizations, respectively. Fully relativistic pseudopotentials were used to incorporate spin–orbit coupling in both structural optimizations and electronic structure calculations.
For the investigation of various CDW patterns, 2$\times$2$\times$2 supercells were employed (see the next section for details), and to compare the band dispersion with the angle-resolved photoemission spectra, we utilized the band unfolding technique \cite{ku2010unfolding,Zheng_VaspBandUnfolding}. Space group symmetries were determined from the relaxed structures using the {\sc findsym} package, with a tolerance in lattice and atomic displacements of 10$^{-3}$\AA ~\cite{Stokes2005Findsym,findsym}.

For the ARPES measurements, 1\textit{T}-TiSe$_2$ single crystals were purchased from HQ Graphene. ARPES data were taken at the HERS endstation of Beamline 10.0.1 at the Advanced Light Source, Lawrence Berkeley National Laboratory, using a Scienta R4000 analyzer. The base pressure was better than 4 $\times$ 10$^{-11}$ Torr.
The photon energy was set for $p$-polarization with energy and angular resolutions of 18–25 meV and 0.1°, respectively. To achieve high-quality ARPES data for bulk 1\textit{T}-TiSe$_2$, the samples were cleaved in ultra-high vacuum at 20 K.

The temperature dependence of the electrical resistivity was measured on single-crystalline 1T-TiSe$_2$ sample using a TeslatronPT (Oxford Instrument, 12 T magnet system) at the Quantum Matter Core Facility of Pusan National University. Longitudinal resistivity (ρ$_x$$_x$) was recorded using a standard four-probe method over the temperature range of 4−300 K. Electrical contacts were made using gold wires and silver paste. The constant current 5 mA was applied, and the voltage was measured using a Lakeshore M81 lock-in amplifier at 13.333 Hz (Quantum Matter Core Facility of Pusan National University) to reduce the noise and improve the signal. The measurement was performed at ambient pressure (1 atm) with a temperature sweep rate of 0.8 K/min. The measured resistivity data is consistent with previously reported ones (see Fig.~S1 in the Supplementary Information)\cite{ueda2021correlation}.

\section{\label{sec:level3} Results}
\subsection{\label{sec:level3} Charge density wave orders predicted from calculations}

%{\color{violet}\bf (결과 들어가기 전에, TiSe2 low-temperature CDW 구조 및 space group symmetry에 대한 기존 결과들 정리가 필요합니다. 논문 서치 및 정리해서 여기 언급 및 cite해 주세요. 그래야 우리가 우리 결과를 이야기 시작할 때, 우리 연구가 기존 연구들과 다른 점을 이야기해 줄 수 있습니다. 이런 식으로 우리 결과를 이야기 시작하는 것이 쉽고 자연스럽습니다.)}
\begin{figure*}[htbp!]
    \includegraphics[width=1\linewidth]{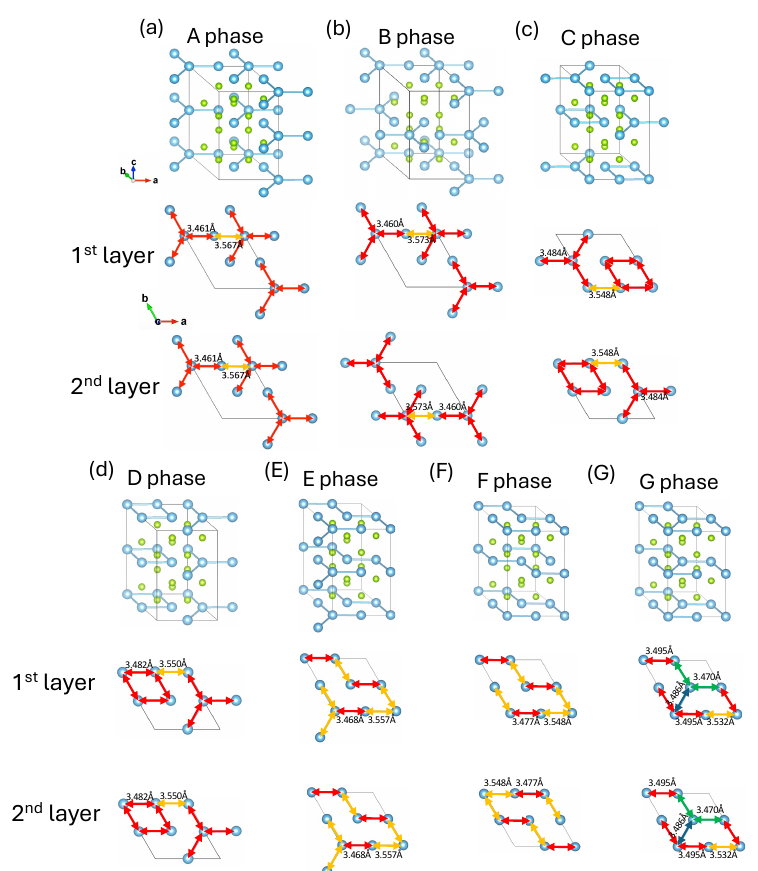}
    \caption{(a–g) Relaxed atomic structures of each phase, with Ti–Ti bonding highlighted for interatomic distances below 3.5~\AA. The A and B phases exhibit Y-shaped Ti–Ti bonding motifs, while the C and D phases show alternating diamond- and Y-shaped patterns. In contrast, the E, F, and G phases are characterized by W-shaped bonding geometries. Numbers are taken from the optB86b-optimized structures.}
    \label{fig:CDWs}
\end{figure*}

\begin{table*}[t]
\caption{Space group symmetries, total energy differences per formula unit, and lattice parameters (from the PBE+optB86b results) of seven structural variants. Note that the 2$\times$2$\times$2 hexagonal supercell settings (not the conventional unit cell scheme) were chosen for the monoclinic structures for the comparison with the trigonal one (Phase B). 
%({\color{violet}\bf 승록씨, 여기 structural parametere들이 실제로 optB86b 계산에서 나왔는지 확인 부탁하고, 나머지 functional 사용한 structural relaxation에서 얻은 lattice parameter들이 optB86b 결과와 크게 다르지 않은지도 확인 부탁합니다. appendix에 다른 functional 들을 통해 얻은 parameter들 넣어야 할 수도 있겠습니다.}
%{\color{orange}\bf 1. 표에있는 lattice parameter 실제 opt86b로 relax한 구조(A,B,C,D)에 TS, DFT-D2로 relax 한 후 또 다시 opt86b로 relax했을때의 구조에 대한것입니다.
%2. 처음 opt86b로 relax를 진행한 후 TS-DFT-D2로 relax하면 새로운 phase가 나왔습니다.})
}

\label{tab:lattice_params}
\begin{ruledtabular}
\begin{tabular}{c|ccccccc}
& Phase A & Phase B & Phase C & Phase D & Phase E & Phase F &  Phase G \\ \hline
\textbf{\begin{tabular}[c]{@{}c@{}}Space\\group\end{tabular}} & $C2$ & $P\bar{3}c1$ & $C2/m$ & $C2$ & $C2$ & $P2/c$ & $C2$ \\
\hline
\textbf{\begin{tabular}[c]{@{}c@{}}opt86b\\$\Delta E$ (meV/f.u.)\end{tabular}} & 0 & 0.25 & –0.30 & 0.03 & 0.13 & 1.41 & 1.16 \\
\hline
\textbf{\begin{tabular}[c]{@{}c@{}}DFT-D2\\$\Delta E$ (meV/f.u.)\end{tabular}} & 0 & -0.07 & –0.10 & 0.34 & -0.28 & -0.87 & -0.39 \\
\hline
\textbf{\begin{tabular}[c]{@{}c@{}}Tkatchenko-Scheffler\\$\Delta E$ (meV/f.u.)\end{tabular}} & 0 & -0.07 & –0.10 & 0.34 & -0.28 & -0.87 & -0.39 \\
\hline
\textbf{\begin{tabular}[c]{@{}c@{}}r$^2$SCAN + rVV10\\$\Delta E$ (meV/f.u.)\end{tabular}} & 0 & -0.65 & –0.68 & 0.26 & 0.52 & 0.56 & 0.53\\ \hline
\textbf{\begin{tabular}[c]{@{}c@{}}Structure\\parameters\\(from optB86b)\end{tabular}} & 
\begin{tabular}[c]{@{}c@{}}a = 7.0290 \AA\\c = 12.0120 \AA\\ $\alpha$ = 89.9710$^{\circ}$\\$\beta$ = 90.0309$^{\circ}$\\$\gamma$ = 120.0220$^{\circ}$\end{tabular} &
\begin{tabular}[c]{@{}c@{}}a = 7.0268 \AA\\c = 12.0110 \AA\\ $\alpha$ = 90.0000$^{\circ}$\\$\beta$ = 90.0000$^{\circ}$\\$\gamma$ = 120.0000$^{\circ}$\end{tabular} &
\begin{tabular}[c]{@{}c@{}}a = 7.0257 \AA\\c = 12.0090 \AA\\ $\alpha$ = 90.0350$^{\circ}$\\$\beta$ = 90.0330$^{\circ}$\\$\gamma$ = 119.9680$^{\circ}$\end{tabular} &
\begin{tabular}[c]{@{}c@{}}a = 7.0270 \AA\\c = 12.0120 \AA\\ $\alpha$ = 89.9710$^{\circ}$\\$\beta$ = 90.0309$^{\circ}$\\$\gamma$ = 119.9730$^{\circ}$\end{tabular} &
\begin{tabular}[c]{@{}c@{}}a = 7.0257 \AA\\c = 12.0298 \AA\\ $\alpha$ = 89.9861$^{\circ}$\\$\beta$ = 90.0139$^{\circ}$\\$\gamma$ = 120.0214$^{\circ}$\end{tabular} &
\begin{tabular}[c]{@{}c@{}}a = 7.0250 \AA\\c = 12.0283 \AA\\ $\alpha$ = 89.9696$^{\circ}$\\$\beta$ = 90.0304$^{\circ}$\\$\gamma$ = 120.0557$^{\circ}$\end{tabular} &
\begin{tabular}[c]{@{}c@{}}a = 7.0255 \AA\\c = 12.0286 \AA\\ $\alpha$ = 89.9806$^{\circ}$\\$\beta$ = 90.0194$^{\circ}$\\$\gamma$ = 120.0248$^{\circ}$\end{tabular} 
\end{tabular}
\end{ruledtabular}
\end{table*}

%These conflicting results highlight the complexity of determining the exact nature of the ground-state CDW pattern in TiSe$_2$. 
%This disagreement has led to various interpretations regarding the structural phase and its associated space group symmetry, with different experimental techniques yielding distinct findings.

%{\color{red}
%Figure \ref{fig:disp}(a) shows the projected band structure of 1\textit{T}-TiSe$_2$. In 1\textit{T}-TiSe$_2$, the conduction band is mainly characterized by the Ti d orbitals, while the valence band is mainly dominated by the p orbitals of the Ti and Se atoms. Figure \ref{fig:disp}(b) illustrates the atomic displacements associated with the CDW, as reported in previous studies. Owing to phonon instability at the M and L points, Ti atoms shift in the direction indicated by red arrows, whereas Se atoms undergo displacement along the blue arrows directions.
%}
%
%{\color{violet}\bf (그림 1 캡션에 제가 남긴 코멘트를 참고 부탁드립니다. 우리가 이 논문에서 밴드의 orbital character를 자세히 들여다 보지는 않고, 구조 변화 및 이에 따른 band 모양 변화 자체에 초점을 맞추고 있기 때문에, CDW 있기 전의 band orbital character에 대해 이야기할 부분이 별로 없습니다. 따라서 여기서는 phonon band 및 여기서 보이는 soft mode들 설명에 초점을 맞추는 것이 더 좋아 보입니다. 다만 phonon DOS는 크게 쓸 데가 없는데, 포함해도 무방합니다. 다만 phonon band 옆에 작게 붙이는 느낌으로...)}
%
\begin{figure*}
    \includegraphics[width=1\linewidth]{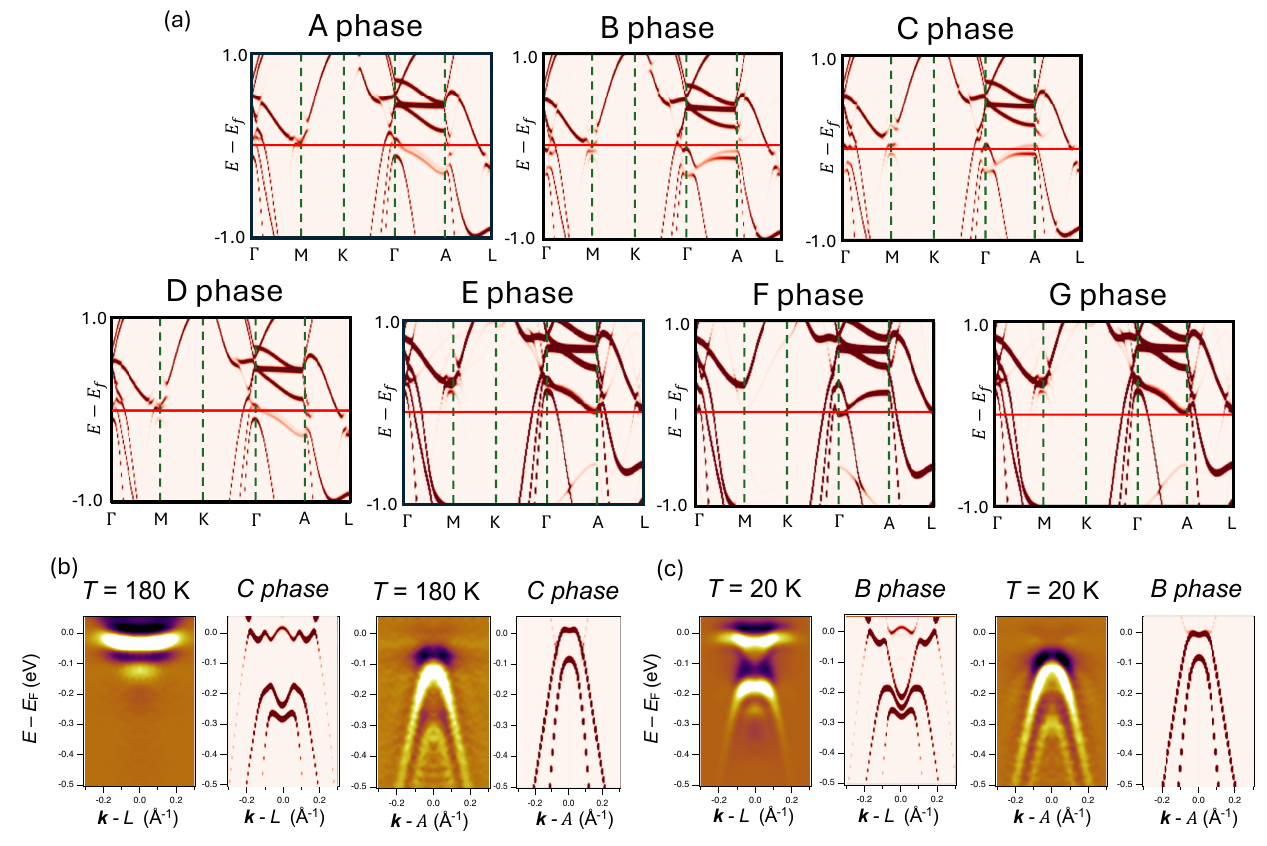}
    \caption{(a) DFT band structure of each phases. (b) Left: ARPES data (using 75 eV photon) at 180 K and 20 K for the L and A points. Right: DFT band structure for the A and C phases at the L and A points.}
    \label{fig:unfolded}
\end{figure*}

\begin{figure}[h] % 'h'는 현재 위치에 최대한 배치
    \flushleft % 그림을 왼쪽 정렬
    \includegraphics[width=0.5\textwidth]{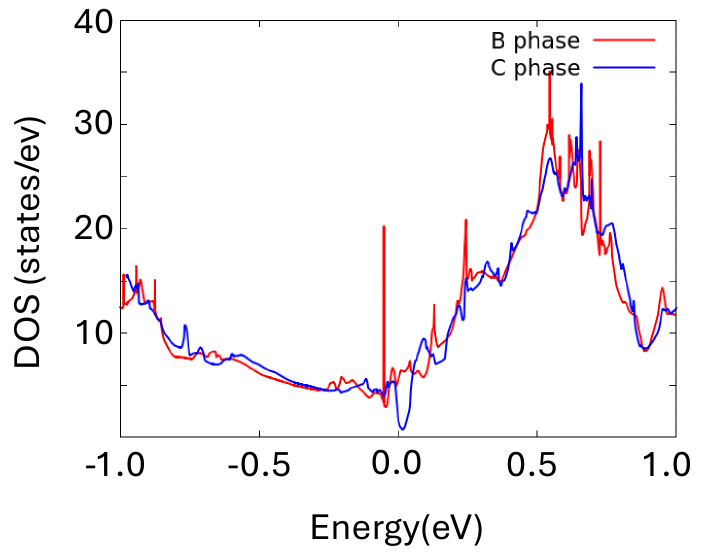} % 이미지 삽입 (파일명 변경 필요)
    \caption{Density of states for the B and C phases. The C phase exhibits a dip near the Fermi level, whereas the B phase does not.
    }
    \label{fig:dos}
\end{figure}

%The nature of the low-temperature CDW structure of TiSe$_2$, specifically whether the space group is $P\bar{3}c1$ \cite{ueda2021correlation,hellgren2017critical} or $C2$ \cite{kim2024origin,wickramaratne2022photoinduced}, remains a topic of ongoing debate. 
To explore various types of CDW that occur beyond the harmonic approximation, we employed 2$\times$2$\times$2 supercells by initially introducing arbitrary Ti atomic displacement patterns ($\delta R_{\rm Ti} = 0.14$\AA) and performing full structural relaxations afterward without any symmetry constraint. Consequently, seven distinct CDW structures were identified.
Figure~\ref{fig:CDWs}(a–g) presents the optimized structures of these seven structural phases, and Table~\ref{tab:lattice_params} lists the lattice parameters, space group symmetries, and total energies relative to that of the A phase. 
The A phase corresponds to a 2$\times$2$\times$1 CDW structure, featuring the formation of Y-shaped Ti clusters\cite{Guster2018} (see Fig.~\ref{fig:CDWs}(a)). When two inverted A-type layers are alternatively stacked (namely the stacking of Y-shaped and inverted-Y-shaped Ti clusters, see Fig.~\ref{fig:CDWs}(b)), it becomes the B phase. Note that the B phase has been proposed as the ground state of 1$T$-TiSe$_2$ in previous studies\cite{DiSalvo1976,ueda2021correlation}, which was also found to be a possible metastable state from a previous density functional theory study\cite{subedi2022trigonal}. Interestingly, while the space group symmetry of the B phase is $P\bar{3}c1$, retaining the three-fold symmetry, the A phase shows a small but finite monoclinic tilting of the unit cell and has a $C2$ symmetry (see Table~\ref{tab:lattice_params}).

A similar stacking relationship is observed between the C and D phases (Fig.~\ref{fig:CDWs}(c-d)), where the formation of in-plane alternating Y- and diamond-shaped Ti clusters is found, and the latter exhibits a 180° rotation of the second layer with respect to the former.
 The simultaneous occurrences of both $k_z = 0$- and $\pi/c$-type stacking patterns with the same in-plane CDW order (like Phases A-B and C-D) can be well understood from the appearance of phonon soft modes at both $M$ and $L$ points in the parental $P\bar{3}m1$ structure\cite{Calandra2011}.  
In contrast, the E, F, and G phases exhibit more complex Ti–Ti bonding geometries characterized by W-shaped motifs. Notably, the E phase favors a Y-shaped configuration (Fig.~\ref{fig:CDWs}(e)), the F phase exhibits an L-shaped pattern (Fig.~\ref{fig:CDWs}(f)), and the G phase forms a composite network that bridges W- and L-shaped motifs across layers (Fig.~\ref{fig:CDWs}(g)). We comment that, except Phase A and B, no other phase has been proposed in previous studies.

%{\color{violet}\bf (여기서, 이전 논문들에서 이야기한 구조가 무엇인지, 그리고 우리 결과로 새롭게 밝혀진 CDW 패턴은 무엇인지 이전 논문들 인용하면서 이야기해 줘야 합니다. 예를 들어 Hellgren PRL 2017 논문은 B-phase를 low-T ground state로, 다른 논문들은 A나 D같은 $C2$ symmetry를 low-T phase symmetry로 이야기하고 있지요.)}

%All phases exhibit Y-shaped Ti–Ti bonding, whereas the C and D phases additionally display parallelogram-shaped arrangements. Table 1 summarizes the lattice parameters and space groups of each phase, 

%The A and C phases exhibits the lowest symmetry, classified as C2, while the B phase belongs to the p-3c1 space group. The C assigned to C2/m.
%The observed CDW patterns suggest that a single layer exhibits two distinct types of CDW, and the manner in which adjacent layers stacking whether aligned or staggered gives rise to the four unique 2$\times$2$\times$2 CDW structures.

%{\color{violet}\bf (HSK:왜 A phase가 대칭성이 C2지요? Ti CDW 움직인 모양을 보면 3축 회전 대칭이 성립하는 듯 한데? FindSym에서 tolerance factor 를 10$^{-3}$ angstrom 정도로 주고 space group 한 번 확인해 보세요. 실제 대칭성을 갖는 물질이라도 계산상의 오차 때문에 위치가 원래 있어야 할 곳보다 10$^{-4}$ angstrom 정도는 충분히 틀어질 수 있는데, 이 정도 차이는 무시해야 하는 것이 맞는 듯 합니다.)}
%{\color{orange}\bf findsym으로 확인해본 결과, 위에 테이블과 같은 결과가 나왔습니다.}

We find that (1) the lattice parameters of different CDW phases are almost identical to each other, and (2) so are the energies of different CDW phases, with energy differences being less than 1.41 meV per formula unit (see Table~\ref{tab:lattice_params}).
Our choices of all exchange–correlation functionals (PBE, r$^2$SCAN) and van der Waals corrections (optB86b, DFT-D2, Tkatchenko–Scheffler, and rVV10) consistently reproduce the negligible energy differences between different CDW phases, although the lowest-energy state may vary depending on the functional choice. Note that this quasidegeneracy among different CDW phases is also consistent with a previously reported result\cite{subedi2022trigonal}.
%{\color{blue} (HSK: Here a comment on the triple-q and single-q patterns? For example, E-G phases resembles chain-like formation of Ti atoms which may result in single-q type CDW pattern. But domain effects may mix different phases?)}

Since the occurrence of all seven phases does not involve any significant structural changes such as large differences in the in-plane lattice parameters or layer sliding \cite{haines2018pressure,Harms2020Piezochromism}, it is likely that different types of CDW stacking patterns may stabilize depending on the synthesis conditions in experiments. For example, Ref.~\onlinecite{wickramaratne2022photoinduced} reported the symmetry of the ground state phase of 1\textit{T}-TiSe$_2$ as $C2$, while Ref.~\onlinecite{ueda2021correlation} suggested it to be $P\bar{3}c1$. In addition, a recent elaborate spectroscopic study suggests an exotic chiral CDW state with $P1$ symmetry\cite{kim2024origin}. It should be emphasized that, based on our calculations, that the symmetries of the A, D, E, and G phases are $C2$, signifying the loss of threefold rotation and mirror symmetries even at the single sheet level. In contrast, the B phase exhibits $P\bar{3}c1$ symmetry, indicating the presence of inversion symmetry. Therefore, for example, an alternative stacking of $C2$ layers and $P\bar{3}c1$ layers with different orientations of the twofold axes may lead to the loss of the overall twofold symmetry and the stabilization of the $P1$ structure, providing a way to reconcile the apparently conflicting reports on the ground state symmetry of this compound.

Lastly, we comment that due to the small energy differences among all seven metastable phases, it is difficult to identify the true ground state of 1\textit{T}-TiSe$_2$ in the low-temperature limit based solely on electronic structure calculations. 
In addition, our simulations of orientation-averaged (powder) X-ray diffraction patterns for all seven CDW phases show only negligible differences, making it extremely difficult to distinguish between the various possible CDW phases using diffraction techniques—especially when domains with different CDW orientations coexist, as is likely the case in real samples (see Supplementary Information for further details). 
Therefore, in the following, we employ an alternative strategy of comparing our calculated electronic structures with those obtained from angle-resolved photoemission spectroscopy.

%To assess the feasibility of these CDW phases, we compared their electronic energies, as summarized in Table 1. Given that the A phase corresponds to the previously reported CDW structure, the relative energy differences were evaluated with respect to this reference.
%The energy difference between the lowest and highest phases was approximately 0.277 meV, indicating that all configurations are nearly degenerate. Due to the challenge of distinguishing these phases based solely on energy, we further compared the results from DFT band structure with ARPES data for a more comprehensive analysis.

\subsection{Comparison with angle-resolved photoemission spectra}

For the comparison between the experimental and theoretical electronic structures, we used a band unfolding technique (see the Methods section for further details). Figure~\ref{fig:unfolded}(a) presents the unfolded band structures for each phase. From the comparison between the theoretical and experimental spectra, as shown in Fig.~\ref{fig:unfolded}(b), we conclude that the C and B phases correspond to the CDW patterns above and below 165 K, respectively (see Appendix A for details). Note that the B and C phases are the lowest-energy states in our r$^2$SCAN+rVV10 results (see Table~\ref{tab:lattice_params} for details).

To assess which CDW phase best captures the experimentally observed band structure of 1\textit{T}-TiSe$_2$, we compare our calculated electronic structures with ARPES spectra at representative temperatures. Figure~\ref{fig:unfolded}(b) shows that, at 180 K, the ARPES data at the A point exhibit a well-defined parabolic valence band, a feature that is present in both the B and C phases. However, at the L point, the conduction band appears relatively flat, accompanied by a CDW-induced gap in the valence band. This band topology closely resembles that of the C phase band structure.
At 20 K, the spectral features evolve. While the A point continues to show a parabolic valence band—remaining largely insensitive to temperature changes—the L point undergoes a distinct transformation: the previously flat conduction band sharpens into a V-shaped dispersion, with no clear gap in the valence band. This profile aligns most closely with the electronic structure of the B phase, indicating a crossover in the dominant stacking pattern across the CDW transition.
These observations suggest that different CDW phases may stabilize depending on temperature, with the C phase being favored near the CDW onset and the B phase emerging as the dominant configuration at lower temperatures. 
%
%The marginal differences in total energy ($\sim$ 0.6 meV/f.u.) between the competing phases implies that such a crossover could occur without requiring major structural rearrangements, and may instead reflect a subtle reorganization of the CDW stacking order. 
%
This highlights the utility of ARPES as a sensitive probe of not only band dispersion but also phase identifications in complex CDW systems\cite{Wang2020NC}.
Supporting evidence for the transition from the C to B phases can be found by comparing the computed density of states (DOS) with the temperature-dependent resistivity data.(see Supplymentary Figure S1) Figure~\ref{fig:dos} shows the DOS of the B and C phases, where the latter shows a pseudogap feature very close to the Fermi level. The reduced DOS near the Fermi level in the C phase (compared to that of the B phase) is attributed to the difference in CDW stacking patterns, which can also be seen in the band dispersions near the L point in Fig.~\ref{fig:unfolded}(a–c). The pseudogap feature at the Fermi level in the C phase may lead to semiconducting-like resistivity, while the enhanced DOS in the B phase should result in metallic character below 165 K.

%Ref.\cite{zhang2018unveiling} observed that the pseudogap in 1T-TiSe$_2$ is linked to electronic inhomogeneity, where regions with suppressed CDW gaps exhibit a more pronounced pseudogap feature. This inhomogeneity, driven by local distortions and electron doping, likely contributes to the spatial variation in the pseudogap. It is possible that such fluctuations in the electronic structure, potentially related to metastates, could explain the differences in pseudogap behavior between the C and B phases.

\section{Discussion}

In this work, based on our electronic-structure calculations, we address two unresolved issues in 1\textit{T}-TiSe$_2$: (i) the conflicting reports on the ground-state space-group symmetry and (ii) the origin of the resistivity peak near 165~K. By demonstrating the presence of multiple \textit{quasidegenerate} CDW states, we propose that these issues are closely connected. Nearly degenerate CDW configurations can mix to yield phases with triclinic and/or chiral character, and crossovers between distinct CDW stackings with different electronic structures can naturally account for the observed change in resistivity. The spatial inhomogeneity or anisotropy in CDW patterns reported recently in TiSe$_2$ \cite{zhang2018unveiling,wang2025,guo2025} can likewise be understood as a manifestation of this quasidegenerate landscape.

A natural question is whether more advanced electronic-structure methods (e.g., GW or hybrid functionals) are required to fully determine the low-energy electronic structure and, by extension, the microscopic origin of the CDW (electron–phonon coupling versus exciton condensation) \cite{cazzaniga2012ab, hellgren2021electronic, Pashov2025,yin2024efficient}. Indeed, Ref.~\onlinecite{hellgren2017critical} indicates that introducing nonlocal exchange (HSE) enhances electron–phonon coupling and opens a gap, while subsequent work shows that meta-GGA functionals can partially capture nonlocal screening and reproduce HSE06-level trends \cite{yin2024efficient}. In this light, the small energy differences we find among different CDW phases within PBE and meta-GGA r$^2$SCAN suggest that the conclusion of quasidegeneracy in 1\textit{T}-TiSe$_2$ is \textit{robust} with respect to reasonable upgrades of the exchange–correlation treatment.

A related question concerns the mixing of distinct CDW phases to produce chiral structures or spatially inhomogeneous CDW textures. While the conventional picture assigns a $P\bar{3}c1$ structure at low temperature \cite{DiSalvo1976,ueda2021correlation}, there is also evidence suggestive of chirality \cite{Xu2020,kim2024origin}. Our calculations identify $C2$ structures (Table~\ref{tab:lattice_params}) that are chiral, consistent with other DFT reports \cite{subedi2022trigonal,kim2024microscopic}. As proposed in Ref.~\onlinecite{kim2024origin}, frustration between the lattice and the electronic CDW tendency can yield several nearly degenerate orders—precisely what we find—which in turn can promote the CDW inhomogeneity observed in area-selective electron diffraction \cite{wang2025}. Consideration of \textit{dynamic} CDW fluctuations \cite{Park2023condensation} and their consequences for the electronic structure may therefore be essential to resolving remaining discrepancies \cite{Pashov2025}.

Note that, our observation of vanishingly small energy differences between distinct CDW phases may appear at odds with the well-defined transitions and crossover behavior around 150–200 K. We speculate that, because 1\textit{T}-TiSe$_2$ lies in the strong-coupling regime—characterized by $2\Delta(0)/k_B T_{\rm CDW}\approx 8.7$, where $\Delta(0)$ is the CDW gap\cite{Li2007PRL,guo2025}$,$—the formation of multiple competing domains and the concomitant proliferation of metallic domain walls are energetically disfavored. Another possibility is that long-range Coulomb interactions—relevant to exciton formation or electron–phonon coupling—favor larger characteristic length scales, thereby suppressing fluctuations between different CDW phases.

Lastly, our results motivate targeted nano-probe experiments capable of discriminating among quasidegenerate stackings and quantifying their fluctuations, for example nano-ARPES or nonlinear nanoscopy measurements, for identifying formation of various CDW domains and domain walls, as well as their impacts on electronic structures. 
From a device-engineering perspective, the shallow energy landscape implies that modest external controls—electrostatic gating, uniaxial/biaxial strain, interfacial engineering, or optical driving—should tilt the balance among stackings, enabling reversible and potentially hysteretic switching near the 150–200~K window.

In summary, our combined first-principles calculations and ARPES comparisons clarify the previously ambiguous CDW phenomenology of 1\textit{T}-TiSe$_2$. Identifying multiple quasidegenerate CDW states offers a coherent resolution of the ground-state symmetry debate, while the resistivity peak near 165~K is consistently explained as a crossover between distinct stacking patterns. Beyond resolving long-standing controversies, these findings highlight the necessity of considering nearly degenerate structural configurations when analyzing CDW transitions in layered materials and provide a nano-resolved roadmap—both experimental and conceptual—for controlling CDW order in device-relevant settings.

\begin{acknowledgments}
We thank Bumjoon Kim, Igor Di Marco and Dhani Nafday for fruitful discussions. This work was supported by the Basic Science Research Program through the National Research Foundation of Korea funded by the Ministry of Science and ICT [Grant No. NRF-2020R1C1C1005900, RS-2023-00220471, RS-2023-00221154]. The support of computational resources including technical assistance from the National Supercomputing Center of Korea [Grant No. KSC-2024-CRE-0322] is also acknowledged.
HSK additionally appreciates the hospitality of the Center for Theoretical Physics of Complex Systems at the Institute of Basic Science, Korea, and of the Asia Pacific Center for Theoretical Physics, Korea, during the completion of this work [APCTP-2025-C01]. CH also acknowledges support from the Korea Basic Science Institute (National Research Facilities and Equipment Center) grant funded by the Ministry of Education [Grant No. RS-2021-NF000587 and RS-2024-00435344].
\end{acknowledgments}

%\appendix
%\bibliography{paper,CDW_origin,CDW_temperature,DFT_method,DFT_structure,unfolding,unfolding_tool,TiS2-stacking,tmdc_intro,PBE}
%\section{\label{app:arpes} Comparison between theoretical and experimental electronic structures}


\newpage
\onecolumngrid

\section*{Supplementary Information }
\maketitle

\renewcommand{\thefigure}{S\arabic{figure}}
\setcounter{figure}{0}   

\subsection*{Comparison between simulated powder X-ray diffraction spectra of different CDW phases}

%Figure~\ref{fig:s_diff_1} shows the semi-log plots of the simulated powder X-ray diffraction patterns of our seven charge density wave (CDW) phases as obtained from our calculations. Note that we compare pXRD patterns because in real samples domains of different orientations of CDW patterns are likely to occur in the same sample, resulting in averaged XRD patterns. Note that all CDW phases show clear superstructure peaks when compared to the primitive $P\bar{3}1m$ phase (not shown). It is remarkable that the A, B, C, and D phases bear close similarity to each other in the pXRD pattern, while showing different space group information (See Table~I in the main text). 

%Figure~\ref{fig:s_diff_2}(a) illustrates this similarity even clearer, where the difference between each phase is defined as the integration of the absolute value of pXRD signal over all angle. Therein it is shown that there are two groups that share similar pXRD spectra, namely (A, B, C, D) and (E, F, G). Note that, the A and D phases show almost indistinguishable pXRD patterns (see Fig.~\ref{fig:s_diff_2}(a)), while their CDW patterns and unfolded band structures show clear difference (compare Fig.~1(a) and (d) in the main text and see Fig.~\ref{fig:s_diff_2}(b)). It is remarkable that two CDW phases with distinct superstructure periodicity ($2 \times 2 \times 1$ and $2 \times 2 \times 2$ for A and D phases, respectively) and electronic structure show the almost same pXRD pattern, where the origin of such behavior is still under investigation.

\begin{figure}
    \centering
    \includegraphics[width=1.0\linewidth]{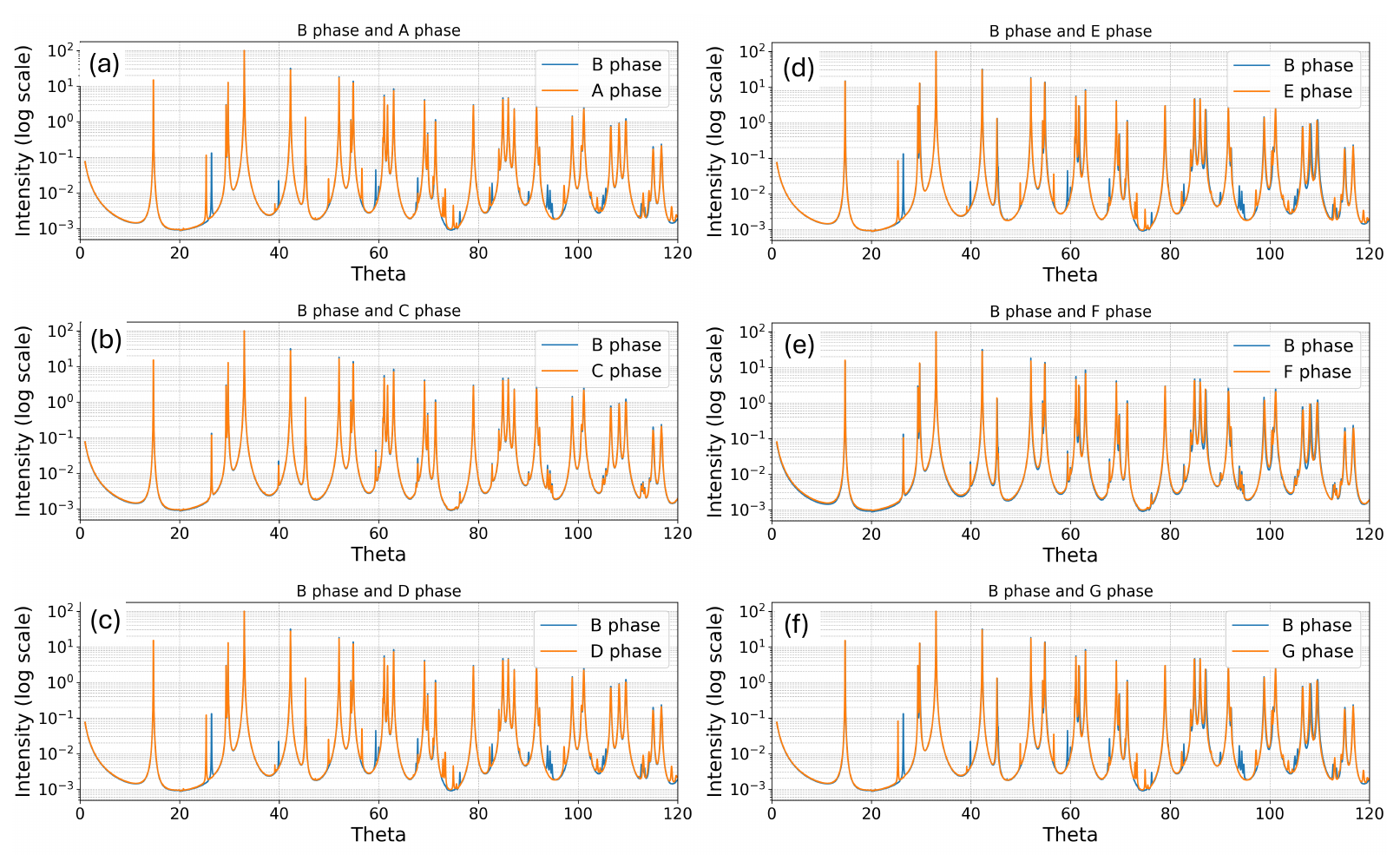}
    \caption{Semi-log plots of simulated powder X-ray diffraction (pXRD) patterns of seven CDW phases obtained from our calculations. Each panel compares the three-fold symmetric B-phase ($P\bar{3}c1$) with other monoclinic (A, C, $\cdots$ G) phases.}
    \label{fig:s_diff_1}
\end{figure}

Figure~\ref{fig:s_diff_1} shows semi-log plots of the simulated powder X-ray diffraction (pXRD) patterns for the seven charge density wave (CDW) phases obtained from our calculations. We focus on pXRD patterns because, in real samples, domains with different orientations of CDW patterns are likely to coexist, resulting in averaged XRD signals. All CDW phases exhibit clear superstructure peaks compared to the primitive $P\bar{3}m1$ phase (not shown). Notably, the A, B, C, and D phases display highly similar pXRD patterns, despite having different space group symmetries (see Table~I in the main text).

\begin{figure}
    \centering
    \includegraphics[width=0.8\linewidth]{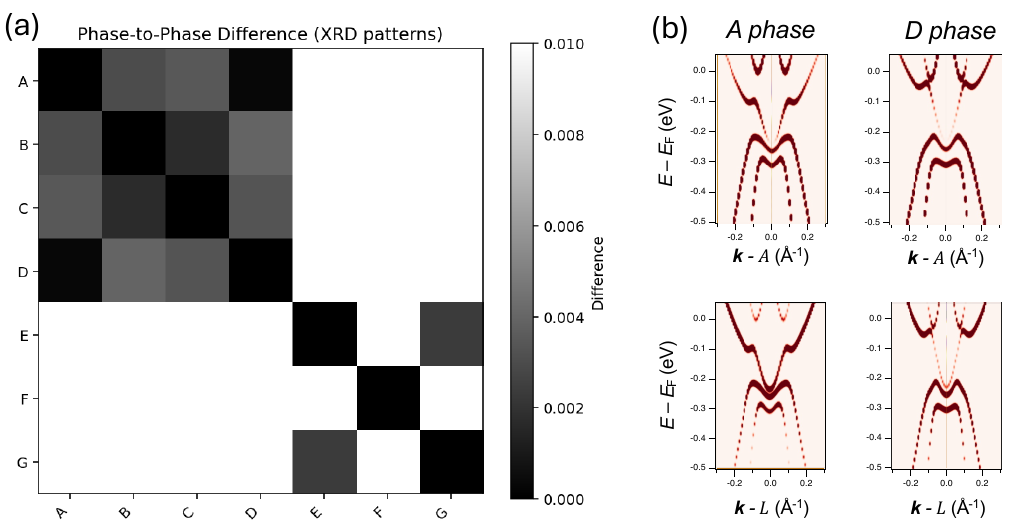}
    \caption{(a) Difference matrix heatmap of the seven CDW phases (in arbitrary unit), where darker (brighter) color indicate more similar (different) powder X-ray diffraction patterns between two structures. (b) Unfolded band spectra of the A and D CDW phases, in the vicinity of A and L points. 
    }
    \label{fig:s_diff_2}
\end{figure}

Figure~\ref{fig:s_diff_2}(a) illustrates this similarity more clearly. Here, the difference between each phase is defined as the integral of the absolute value of the pXRD signal over the entire angle range. The results reveal three distinct groups with similar pXRD spectra: (A, B, C, D), (E, G), and (F). In particular, the A and D phases show nearly indistinguishable pXRD patterns (see Fig.~\ref{fig:s_diff_2}(a)), even though their CDW patterns and unfolded band structures differ significantly (compare Fig.1(a) and 1(d) in the main text, and see Fig.\ref{fig:s_diff_2}(b)). It is striking that two CDW phases with distinct superstructure periodicities—$2 \times 2 \times 1$ for A and $2 \times 2 \times 2$ for D—and different electronic structures yield nearly identical pXRD patterns. The origin of this behavior remains under investigation.

\begin{figure}
    \centering
    \includegraphics[width=0.5\linewidth]{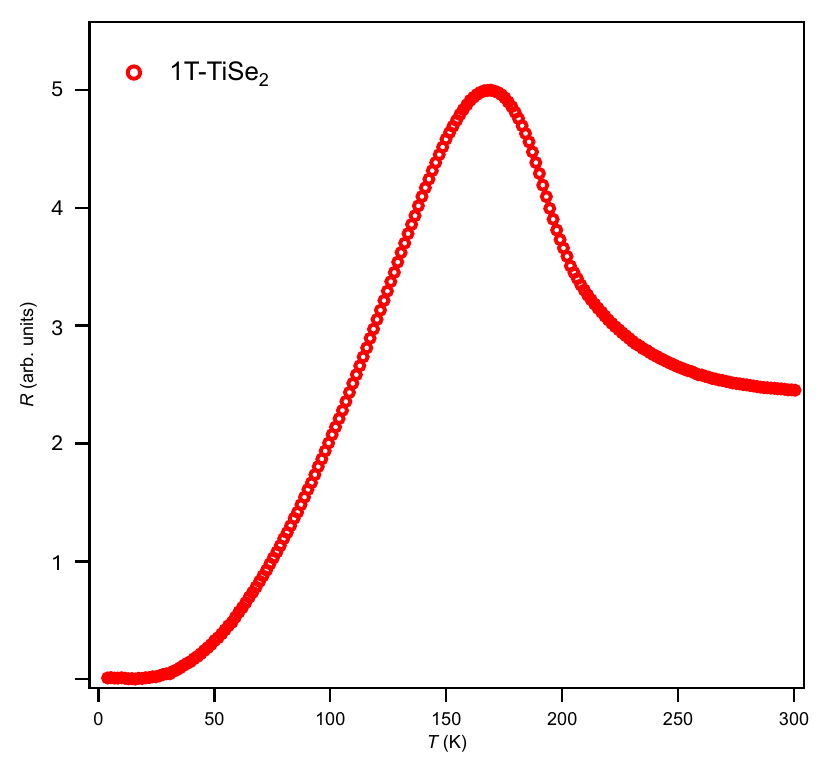}
    \caption{Temperature-dependent resistivity curve of single-crystalline 1T-TiSe$_2$. The measurement was performed over the temperature range of 4–300 K.}
    \label{fig:supplymeny1}
\end{figure}

\subsection*{Temperature-dependent resistivity data}
Figure~\ref{fig:supplymeny1} plots the temperature-dependent resistivity data measured from our sample, which shows the resistivity peak around T = 169 K, consistently with previous observations (H. Ueda {\it et al.}, {\it Phys. Rev. Res.} {\bf 3}, L022003 (2021), etc.). 

\subsection*{Comparison between ARPES and band-unfolding spectra}

\begin{figure*}
    \centering
    \includegraphics[width=1.0\linewidth]{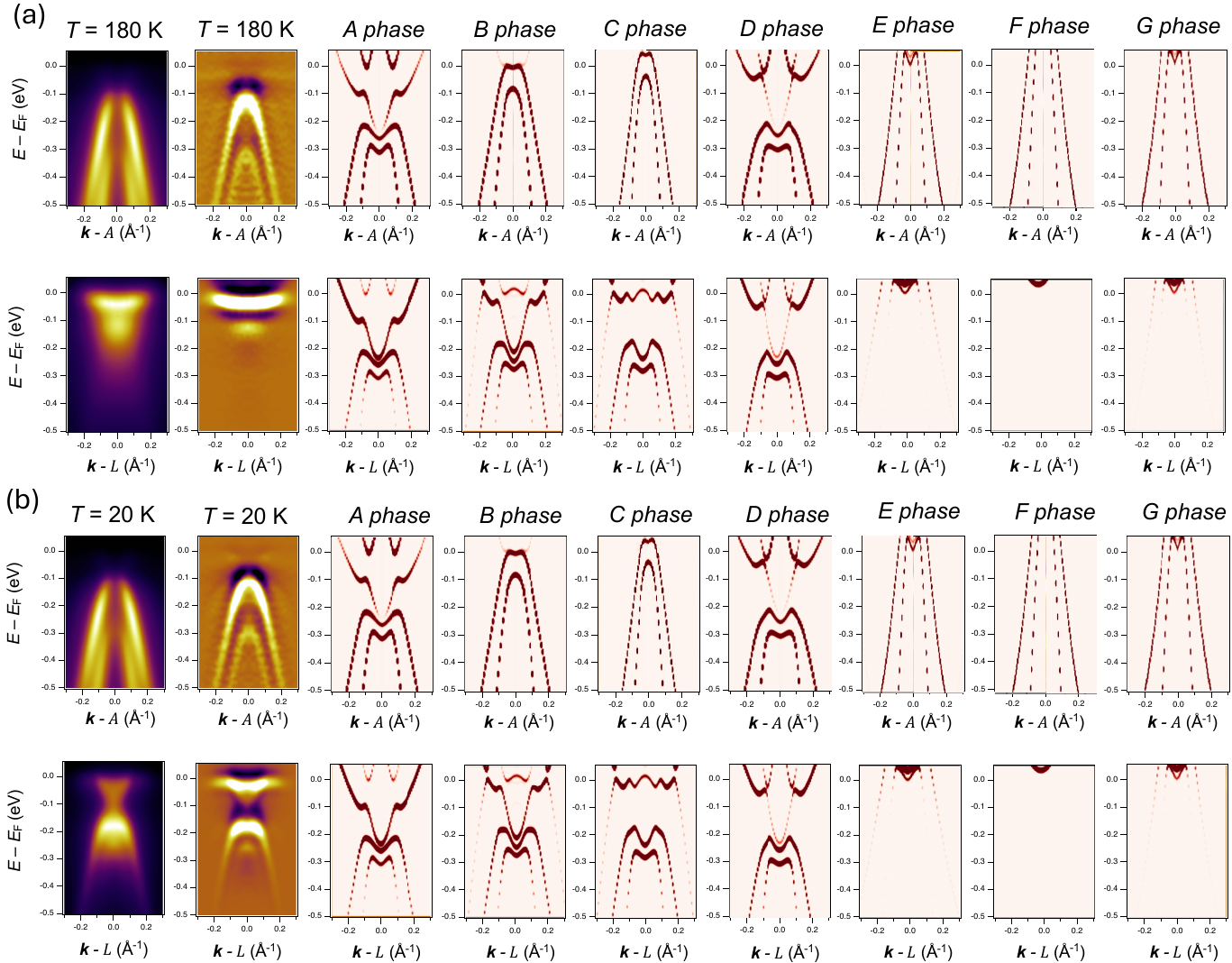}
    \caption{(a, b) ARPES and DFT band structures. Left: Raw and second-derivative ARPES data at the A and L points for 180~K and 20~K, taken using 75 eV photon. Right: DFT band structures for each phase. At 180~K, the ARPES data show a parabolic band at the A point (consistent with the B and C phases) and a flat band at the L point (matching the C phase). At 20~K, below the second phase transition, the ARPES data exhibit a double-dot-like feature at A and a V-shaped band at L, both aligning with the A phase in DFT calculations.}
    \label{fig:supplyment2}
\end{figure*}

Here we compare our ARPES spectra at T = 180 (see Fig.~\ref{fig:supplyment2}(a)) and 20 K (Fig.~\ref{fig:supplyment2}(b)) with band-unfolding spectra of 7 phases (denoted as A, B, C, D, E, F, G phases) from DFT calculations. At 180 K (Fig.~\ref{fig:supplyment2}(a)) an inverted parabolic-like valence bands, with very weak conduction-band-like signatures, are observed from the ARPES data. From the comparison, the B and C phases share similar character, while the A and D phases show M-shaped valence bands around -0.2 eV, unlike the experimental observation. 
%
%a non-parabolic valence band is observed at the A point in the B and C phases, while the A and D phases retain a parabolic valence band. Unlike other phases, which show a V-shaped conduction band at the M point, the D phase uniquely features an open gap.
%
At L point, a flat-band-like signature is observed above -0.2 eV, below which signal is almost absent. The C phase shows similar band shape, while the B phase hosts visible parabolic-like dispersion between -0.3 and -0.1 eV. Based on these observations, the C phase can be considered the best match to the ARPES data at 180 K. 
%
%At 180 K, the ARPES data at the A point exhibit a parabolic band structure, consistent with the B and C phases, while at the L point, a relatively flat band aligns with the C phase. Below the CDW critical temperature, a CDW-induced valence band gap appears at the L point, matching the C phase band structure.
%
At 20 K (Fig.~\ref{fig:supplyment2}(b)), below the second CDW transition temperature, the ARPES data reveal a double-dot-like feature near the Fermi level at the A point, and at the L point a V-shaped conduction band touching valence bands are seen. The inverted parabolic-like valence bands at the A-point match with B and C phases, while the presence of V-shaped conduction-like band above -0.2 eV and the absence of gap between the valence and conduction bands is consistent with the A and B phases. From these, we deduce that the B phase well-matches with the APRES spectra at 20 K, with possibly small fraction of A phase.  
%
%both corresponding to the A phase. Additionally, as the temperature decreases, the ARPES data show a gradual downward shift in the valence bands at both the L and A points. In contrast, the DFT band structures remain unchanged, confirming that these temperature-dependent effects originate from the ARPES measurements.
%
%It is also worth noting that a weak-intensity hole band at the A point, clearly visible in the DFT calculations for the B phase, is not resolved in the ARPES spectra. This missing feature is likely due to matrix element effects or a possible Fermi level mismatch between the experimental conditions and the theoretical calculation.
Note that the mismatch between the ARPES and band-unfolded DFT spectra can be attributed to the absence of matrix element effects in the DFT results, and also to the potential inhomogeneity arising from the mixure of different CDW phases in the sample.

\end{document}